# Testbeds for Transition Metal Dichalcogenide Photonics: Efficacy of Light Emission Enhancement in Monomer vs. Dimer Nanoscale Antennae

Mohammad H. Tahersima[1], M. Danang Birowosuto[3], Zhizhen Ma[1], Ibrahim Sarpkaya[1], William C. Coley[2], Michael D. Valentin[2], I-Hsi Lu[2], Ke Liu[1], Yao Zhou[4], Aimee Martinez[2], Ingrid Liao[2], Brandon N. Davis[2], Joseph Martinez[2], Sahar Naghibi Alvillar[2], Dominic Martinez-Ta[2], Alison Guan[2], Ariana E. Nguyen[2], Cesare Soci[3], Evan Reed[4], Ludwig Bartels[2], Volker J. Sorger[1]

1 Department of Electrical and Computer Engineering, George Washington University, 800 22nd Street NW, Washington, DC 20052, USA

2 Chemistry and Materials Science & Engineering, University of California, Riverside, California 92521, USA

3 Division of Physics and Applied Physics, Nanyang Technological University, 21 Nanyang Link, Singapore 637371, Singapore

4 Department of Materials Science and Engineering, Stanford University, Stanford, California 94305, USA

KEYWORDS: Optical Antenna, Light Emitter, Plasmonic, Florescence, 2D materials, $WS_2$

**ABSTRACT**

Monolayer transition metal dichalcogenides are uniquely-qualified materials for photonics because they combine well-defined tunable direct band gaps and self-passivated surfaces without dangling bonds. However, the atomic thickness of these two-dimensional (2D) materials results in low photo absorption limiting the achievable photo luminescence intensity. Such emission can, in principle, be enhanced via nanoscale antennae resulting in; a) an increased absorption cross-section enhancing pump efficiency, b) an acceleration of the internal emission rate via the Purcell factor mainly by reducing the antenna's optical mode volume beyond the diffraction limit, and c) improved impedance matching of the emitter dipole to the free-space wavelength. Plasmonic dimer antennae show orders of magnitude hot-spot field enhancements when an emitter is positioned exactly at the mid-gap. However, a 2D material cannot be grown, or easily transferred, to reside in mid-gap of the metallic dimer cavity. In addition, a spacer layer between the cavity and the emissive material is required to avoid non-radiative recombination channels. Using both computational and experimental methods, in this work we show that the emission enhancement from a 2D emitter- monomer antenna cavity system rivals that of dimers at much reduced lithographic effort. We rationalize this finding by showing that the emission enhancement in dimer antennae does not specifically originate from the gap of the dimer cavity, but is an average effect originating from the effective cavity crosssection taken below each optical cavity where the emitting 2D film is located. In particular, we test an array of different dimer and monomer antenna geometries and observe a representative ~300% higher emission for both monomer and dimer cavities as compared to intrinsic emission of Chemical Vapor Deposition (CVD)-synthesized $WS_2$ flakes. Observed enhanced light emission from these atomically thin flakes together with the lithographic control of plasmonic antennae on them opens opportunities for engineering light-matter interaction in 2D systems in a test-bed comparable fashion, enabling bright and large-scale 2D opto-electronics.

**INTRODUCTION**

Nanoscale on-chip light-emitting structures are desired in a broad range of fields including displays, sensors, and optical interconnects. Currently, solid-state light-emitting diodes are based on direct band gap compound semiconductors such as gallium arsenide or gallium nitride that require epitaxy for growth posing challenges for integration with silicon technology. Moreover, for nanoscale devices with increased surface to volume ratio, these materials suffer from high parasitic surface velocity recombination rates that limit the internal quantum efficiency [1, 2]. Transition-metal dichalcogenides (TMD) are 2D materials, whose layered structure offers self-passivated surfaces without out-of-plane dangling bonds, and have a direct bandgap when thinned to a monolayer due to the emergence of a direct bangap at the k-space $\Gamma$ point. Here, they offer high quantum-yield and the potential for pure excitonic states [3-5]. As such, TMDs are attractive materials for novel nanoscale optical emitters and optoelectronic devices [3, 7-12].

The direct band gap of monolayers of several chemically, structurally and electronically similar semiconducting TMDs such as $MoS_2$, $MoSe_2$, $WS_2$, and $WSe_2$ has demonstrated light emission in the visible and near-infrared spectral regions. The spatial confinement of carriers to a 3-atom thin physical plane and the weak dielectric screening in atomically-thin materials lead to high oscillator strengths and strong Coulomb interactions between the excited electron in the conduction band and the remaining hole in the valence band. Resulting strong exciton binding energy makes the observation of excitons possible even at the room temperature [13-20]. In addition to neutral excitons, charged trions can also be excited in the presence of residual excess charge carriers. These quasi particles consist either of two electrons and one hole ($A^-$), or one electron and two holes ($A^+$). Electrostatic gating, therefore, modifies the spectral weight of charge-neutral excitonic species in TMDs [16-19]. Moreover, given the large binding energy of the excitons, the formation of states consisting of two excitons (biexciton) is possible in TMDs, whose photoluminescence (PL) emission is red-shifted due to the additional binding energy [16,20].

A drawback of TMD materials for optic and photonic applications is the low modal overlap with any optical field originating from the atomic thickness of the monolayer material and the fundamentally weak light-matter interaction of bosons with fermions. This can be enhanced via resonant (cavity, surface-plasmon-resonance) and non-resonant (waveguide dispersion, metamaterials, index tuning) systems. Antennae fall into the former category and they can synergistically a) increase the absorption cross-section thereby enhancing the pump efficiency, b) accelerate the internal emission rate via the Purcell factor through the nanoscale optical mode of the antenna, and c) improve emission out-coupling to free-space via impedance matching (transformer action). As such, optical antennae increase the excitation rate while simultaneously enhancing the local density of states (DOS) in the emission process, which modifies (here accelerates) the spontaneous emission rate (Purcell effect). Hence, these optical antennae behave as electromagnetic cavities that strongly modify spontaneous emission of fluorescence in the spatial and spectral proximity [22]. Plasmonic antennae are unique cavities; a good antenna has a low quality (Q)-factor and is, thus, an effective radiator. However, the light-matter interaction enhancement quantifier, or Purcell factor, is proportional to the ratio of Q/V, where V is the optical mode volume. Given the possibility for sub-diffraction limited plasmonic optical mode volumes, the inherently low Q factors ensure decent antenna function, while high Purcell factors

~10's to 100's is obtainable [24,34,35]. Compared to photonic high-Q cavities, plasmonic antennae allow for simultaneous high absorption and PL emission. Antenna-enhanced light-emitters have short radiative lifetimes and can have a deep subwavelength optical mode, thus opening the possibility of creating ultrafast, nanoscale emitters [23].

Recently, multiple TMD-plasmonic hybrid nanostructures have been investigated [24-32]. A common claim is that the gap-mode of the plasmonic dimer's field enhancement enables high Purcell factors. Plasmonic dimers, which are nanoscale structures consisting of two metallic nanoparticles separated by a small gap, support hybridized plasmon resonances because of the capacitive coupling between the plasmon modes of each nanoparticle. For a quantum emitter (e.g. <20nm quantum dot) placed inside this gap, the coupling strongly localizes charges at the junction between the two nanoparticles, giving rise to large field enhancements at the center of the feed-gap of the dimer antenna. However, for emitters that are not comparable in size to a quantum dot in all three dimensions, such as a TMD layer, the emitter must be offset from the antenna either below or on top of the antenna and, thus, is unable to take the advantage of the highest density of states (DOS) at the hotspot. For TMD emitters with zero distance from the plasmonic antennae, the strong field gradients of the point dipole source can efficiently excite lossy multipolar modes of the antenna which are mostly dark or weakly coupled to the radiation field and, therefore, convert the electromagnetic energy mostly into heat [31]. To avoid the emitter quenching effect and the coupling to lossy plasmonic surface waves, the emitter should be separated from the metallic nanoparticle by a distance previously reported to be about ~8 to 10 nm [33-35]. Before we tested fabricated TMD-nanocavity systems, we modeled a broad spectrum of dimer and monomer antenna configurations and find that – within the limitations imposed by the 2D geometry of the film and the need for a separator - the maximum field enhancement of a dimer antenna relative to a monomer disk is only about twofold. For such designs, we observe that the pertinent cavity field enhancement in the dimer case does not originate from the gap between the metal particles, but from each monomer disc. We further compare the TMD emission enhancement of monomer vs. dimer antennae relative to intrinsic PL emission.

Our work uses exclusive high-quality $WS_2$ single-layer islands prepared by chemical vapor deposition (CVD). Using material obtained by a scalable deposition technique forwards relevance of our finding for future transition technological implementation and mass production. Prior work [24, 27, 30] on plasmonic antennae utilized material obtained by exfoliation, a non-scalable technique. The CVD process is the preferred method for synthesizing TMD materials due to the pristine monolayer quality of the materials including a high-level of process control as readily demonstrated by the semiconductor industry. Electron beam lithography (EBL) has been used for fine-tuning of the optical antenna dimensions of up to 10's of nanometers precision (see methods).

Although in atomically thin layered TMDs, atoms are confined in a plane, the electric field originating from charges in the 2D crystals have both in-plane as well as out-of-plane components. Moreover, the large surface to volume ratio in 2D materials enhances the significance of surface interactions and charging effects. Thus, the dielectric permittivity mismatch between the 2D semiconductor materials, the surrounding environment, and an induced strain from capping material can intricately affect the electronic and optoelectronic properties of low dimensional materials [38].

In this article, we demonstrate that spontaneous emission of atomically thin $WS_2$ film coupled to 4 different plasmonic nano-cavity design can be substantially enhanced up to 300% compare to intrinsic emission of Chemical Vapor Deposition (CVD)-synthesized $WS_2$ flakes. More importantly, we have shown that fluorescent enhancement of nano antenna coupled 2D materials unlike quantum dots is an areal average effect rather than hot-spot like effect. We also discuss how the Off- and ON-resonance plasmonic pumping of the monolayer $WS_2$ film excitonic luminescence is susceptible to electric field intensity variations caused by surface plasmons.

**MATERIALS**

The sample studied here consists of $WS_2$ monolayer flakes grown directly by CVD on 100 nm of thermal $SiO_2$ on Si wafer. A bright field microscopy image in Fig.1a shows the bare substrate appearing purple, the single-layer material as dark blue and thicker material regions as lighter blue areas. To characterize our emitter material, micro-Raman, micro-PL and differential reflectance spectra were taken from $WS_2$ flakes on a $SiO_2$/Si substrate at the room temperature. We confirm the single-layer character of the $WS_2$ material by the appropriate difference in the intensity of the interlayer phonon mode $A_{1g}$ [39,40] between material identified as multilayer and single-layer (Figure1 b,c) as well as a corresponding difference in PL intensity. We further analyze the Raman signal by a multi-Lorentzian fitting of all recognizable features in both monolayer and few layer $WS_2$ (Figure 1c). The strongest peak at ~354 $cm^{-1}$ is attributed to the in-plane $E_{2g}^1$ and second order vibrational mode 2LA(M), and the peak at ~418 $cm^{-1}$ to the out of plane vibrational mode $A_{1g}$(G). Our measured Raman modes match well with our theoretically (DFT) calculated Raman modes (see methods section for further details). The $A_{1g}$(G) mode blue shifts, and its relative intensity to the in-plane vibrational mode component increases, with an increasing number of $WS_2$ layers. This is expected due to a stronger force among the layers caused by van der Waals interactions [40].

The emission spectrum of $WS_2$ shows a dependence on flake thickness, resulting in a drastic increase in the emission quantum efficiency on monolayer $WS_2$ films indicating the indirect-to-direct bandgap transition upon multi-layer to monolayer scaling (Inset in Figure 1a) [41]. As expected, PL spectra of $WS_2$ monolayers are excitonic in nature, exhibiting strong emission corresponding to *A* excitonic absorption at 633 nm (~1.96 eV).

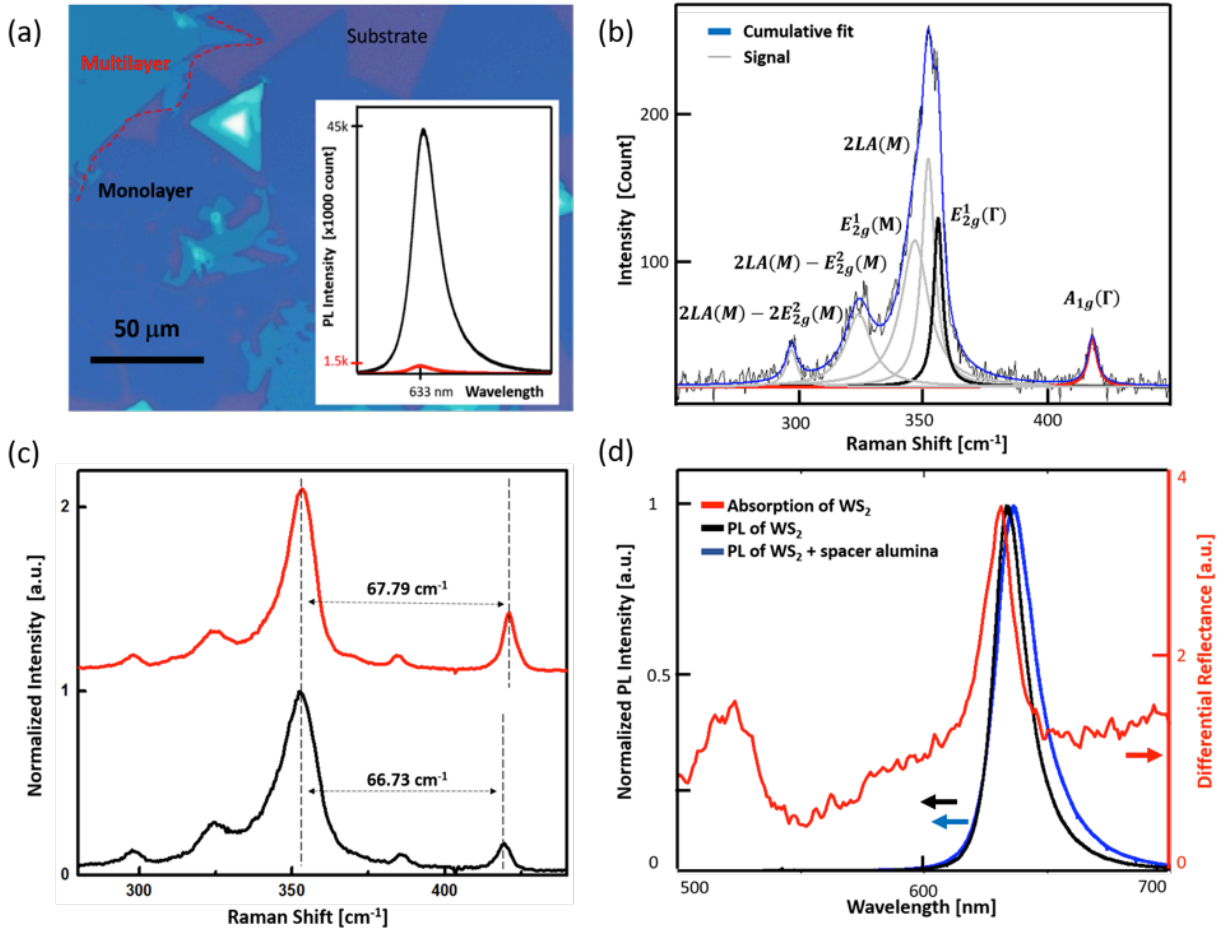

Figure 1. Optical characterization of CVD-grown $WS_2$ at room temperature. (a) Optical images of as-grown $WS_2$ on 100 nm $SiO_2$ on silicon substrate. (inset) Comparison of PL emission of monolayer and multilayer $WS_2$ (b) Room-temperature Raman spectra from a monolayer $WS_2$ flake, including Lorentzian peak fits for the 532 nm laser excitation. (c) Normalized Raman spectra of as-grown monolayer and bilayer $WS_2$. (d) PL (black curve) and absorption (red curve) spectra of as-grown $WS_2$ monolayers. Blue curve shows the PL spectrum of as-grown $WS_2$ after deposition of few nm alumina.

We further measure the differential reflectance of monolayer $WS_2$ to obtain its absorption spectrum in the visible spectrum range, and find the *A* excitonic absorption peak at 631 nm (~1.96 eV). Excitonic emission from monolayer $WS_2$ on $SiO_2$/Si substrate demonstrates a very small stokes shift of ~2 nm (less than 0.01 eV) between positions of the band maxima of the absorption and emission spectra (Figure 1d) like the previous works. [32, 41]. The emission spectrum of monolayer $WS_2$ is relatively narrow with a full width at half the maximum (FWHM) value of about 17 nm. For quantum-well structures, the Stokes shift and FWHM are indicators of interfacial quality. For instance, the magnitude of Stokes shift in monolayer TMD is found to increase with doping concentration [42]. Thus, the narrow emission spectra, whose FWHM is comparable to thermal energy at room temperature along with small Stokes shift, indicates the high quality of $WS_2$ flakes as demonstrated here. The absorption spectrum of monolayer $WS_2$ shows two excitonic absorption peaks *A* and *B*, mainly arising from direct gap transitions at the K point, which are located at 631 nm (1.96 eV) and 522 nm (2.38 eV) respectively (Figure 1d). The energy difference between the *A* and *B* peaks is approximately 420 meV, an indication for spin-orbit coupling in monolayer $WS_2$. We find an additional peak near 460 nm, which arises from the DOS peaks in the valance and conduction bands. However, since we carried out our measurements at room temperature, we do not observe previously reported *A'* and *B'* peaks which are believed to stem from a splitting of the ground and excited states of *A* and *B* transitions [39].

**RESULTS AND DISCUSSION**

The PL emission spectrum of monolayer $WS_2$ flake evolves from a single peak at 633 nm (~1.96 eV) wavelength confirming the PL origin from an *A* exciton. After deposition of the $Al_2O_3$ spacer layer, PL emission spectrum of monolayer $WS_2$ under low excitation powers slightly redshifts from the pristine value of 633 nm (~1.96 eV) to 637 nm (~1.95 eV) (Figure 1d).

We attribute this redshift to the strain imparted by the $Al_2O_3$ during the oxide deposition by means of electron beam evaporation. Following the deposition of the antennae, we find an enhancement of the PL emission of monolayer $WS_2$ flakes for the case of 75 nm dimer cavity by a factor of 3.2 (2.7) in peak intensity (in integrated PL count) relative to emission from the reference sample (bare monolayer $WS_2$ flake) at the same excitation power density (Figure 5). As can be seen from Figure 5a, in some cases such as, monomer and dimer antennae with a 200 nm radius, the PL peak of nano-antenna coupled monolayer $WS_2$ does not substantially change. We note that some fluctuations for the enhancement values and spectral response shifts can be expected, because the optical properties of the $WS_2$ monolayer are strongly influenced by the nanoantenna surface plasmon that can alter the effective pumping of $WS_2$, generation rate of electron hole pairs, and quantum yield of the emitter system, discussed below.

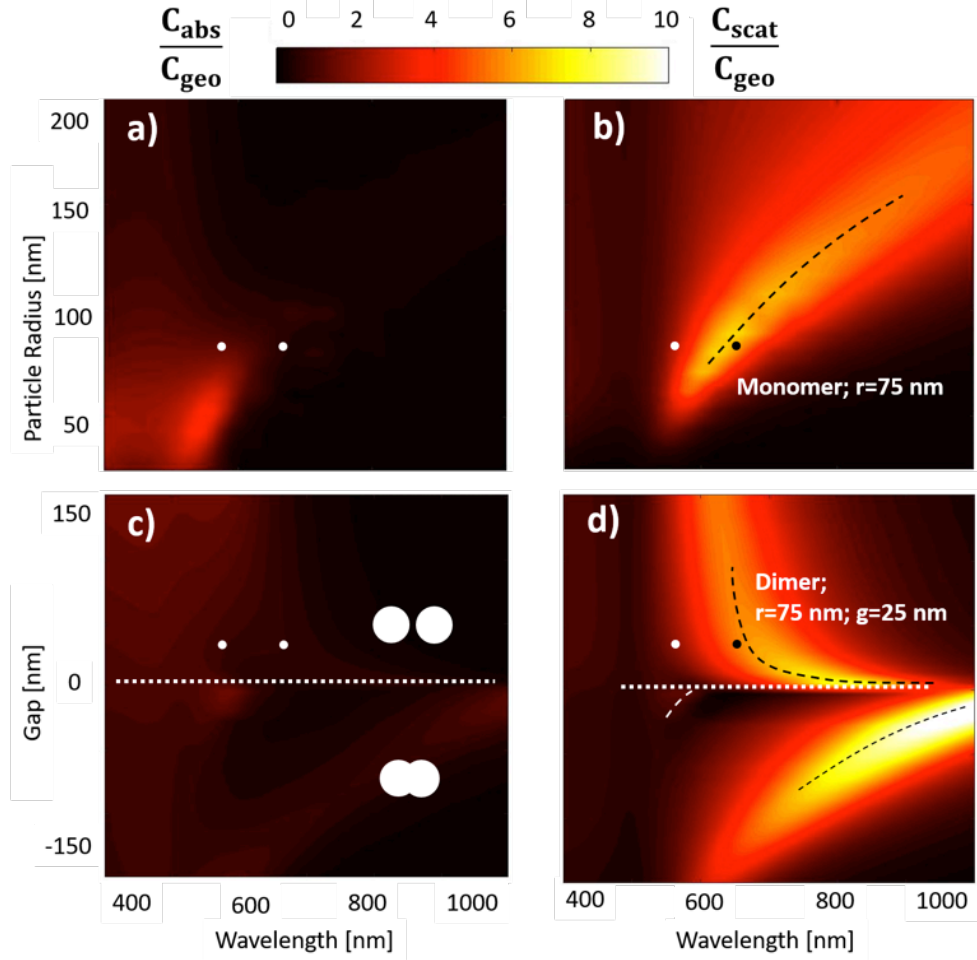

Figure 2. **Cold cavity response.** Absorption loss ($Q_{abs}$) **(a)** and far field scattering efficiency ($Q_{scat}$) **(b)** mapping of monomer nanodisc antennae for radius range of 50 nm to 200 nm (The black and white points represent our fabricated antennae for excitation and emission wavelengths respectively. Absorption loss **(c)** and far field scattering efficiency **(d)** for dimer nanoantenna of single 75 nm radius and in a gap sweeping range from -75 nm (overlapping charge transfer mode) to 75 nm (gap plasmon mode). The points represent our fabricated 75 nm dimer antenna. The scale bar is the ratio of absorption or scattering cross section to geometrical cross section of each type of antenna.

When a quantum emitter interacts with the local fields of an optical antenna, the coupled system has a larger absorption cross-section compared with that of the isolated emitter leading to an optical concentration effect enhancing the effective pump intensity. To reveal such far-field cold-cavity effect and to find the resonances for the monomer and dimer nano-antennae, we analytically describe their spectral scattering efficiency and absorption loss response by dividing both the absorption cross section area and scattering cross section area of each individual antenna by its geometrical area:

$$Q_{ext} = Q_{scat} + Q_{abs} = \frac{C_{ext}}{C_{geo}} = \frac{C_{scat}+C_{abs}}{C_{geo}} \qquad (1)$$

where $Q_{ext}$, $Q_{scat}$, $Q_{abs}$ ($C_{scat}$, $C_{abs}$) are extinction, scattering, and absorption efficiency (cross section), respectively. $C_{geo}$ is the geometrical cross section of the antenna which depends on its radius. The geometrical cross section for a dimer antenna is twice that of a monomer antenna due to its factor of two larger surface area. We note that $Q_{abs}$ is a parasitic part of the cold-cavity and is to be minimized, and it should not be evaluated as the emitter absorption where an enhancement is preferred as long as the system operates in the linear regime. The resonance wavelength and scattering cross section of the monomer antenna is a function of the permittivity of the plasmonic material and the dimension of the antenna. We have chosen gold for fabrication of our antennae due to its Drude-like response for wavelengths above 600 nm. Therefore, the monomer antenna with a 75 nm radius has a scattering efficiency of about 1.8 and 6.3 at excitation wavelength of 532 nm and emission wavelength of ~640 nm, respectively, in which the emission of the $WS_2$ is in resonance with the resonance of the cavity (Figure 2b). Since the monomer is electromagnetically a simple dipole under excitation, we observe the expected monotonic resonance redshift with increasing dimension of the monomer particle, while the discrepancy from a linear trend can be explained by dispersion. The spectral-design areas to avoid absorption losses are a) near the blue frequencies in the visible, and b) small monomers (<75 nm) (Figure 2a).

The dimer antennae fall into two categories depending on whether the interparticle distances (i.e. gap) is positive (true dimer), or negative (lumped dimer) (Figure 2c). For the dimer antennae, the radius of each discs is kept at the constant values of 75 nm, 100 nm, and 200 nm and the interparticle distance in between them is swept from negative to positive values, where the minus values of gap are for overlapping dimer disks. Comparing spectral resonances of these three cases, we find that the lumped dimer effectively behaves as a monomer with a metal particle diameter about equal the total length of the dimer (i.e. 2x diameter-gap) (black dashed line, Figure 3d). Although the scattering efficiency in the dimer antenna is less than the scattering efficiency of its corresponding monomer antenna with same radius, the scattering cross section is larger (almost twice) relative to that of monomer antennae. As expected, the resonance for the large positive gap dimer approaches that of the monomer one.

Moreover, the emission intensity profile of an emitter in an optical-cavity-antenna is governed by:

$$I_{out} = (I_{pump})_{eff} (\eta_{int})(\eta_{oc})$$

$$= \left[I_0 \cdot \left(\frac{\iint \frac{|E|^2}{|E_0|^2} ds}{C_{geo}}\right)\right]_{\lambda-Pump} [F_p \cdot QE]_{\lambda-Emit} \qquad (2)$$

where $I_{out}$ is the outgoing emission from the system, $I_{Pump}$ is the effective incoming laser pump intensity on the monolayer WS$_2$ emitter, $\eta_{int}$ is the internal photon generation process of the emitter (Purcell), $\eta_{oc}$ is the outcoupling efficiency of emitted photons from the monolayer WS$_2$ away via the cavity-antenna system into free space above the sample where the light is collected at the microscope objective, represented here as the quantum efficiency of the antenna.

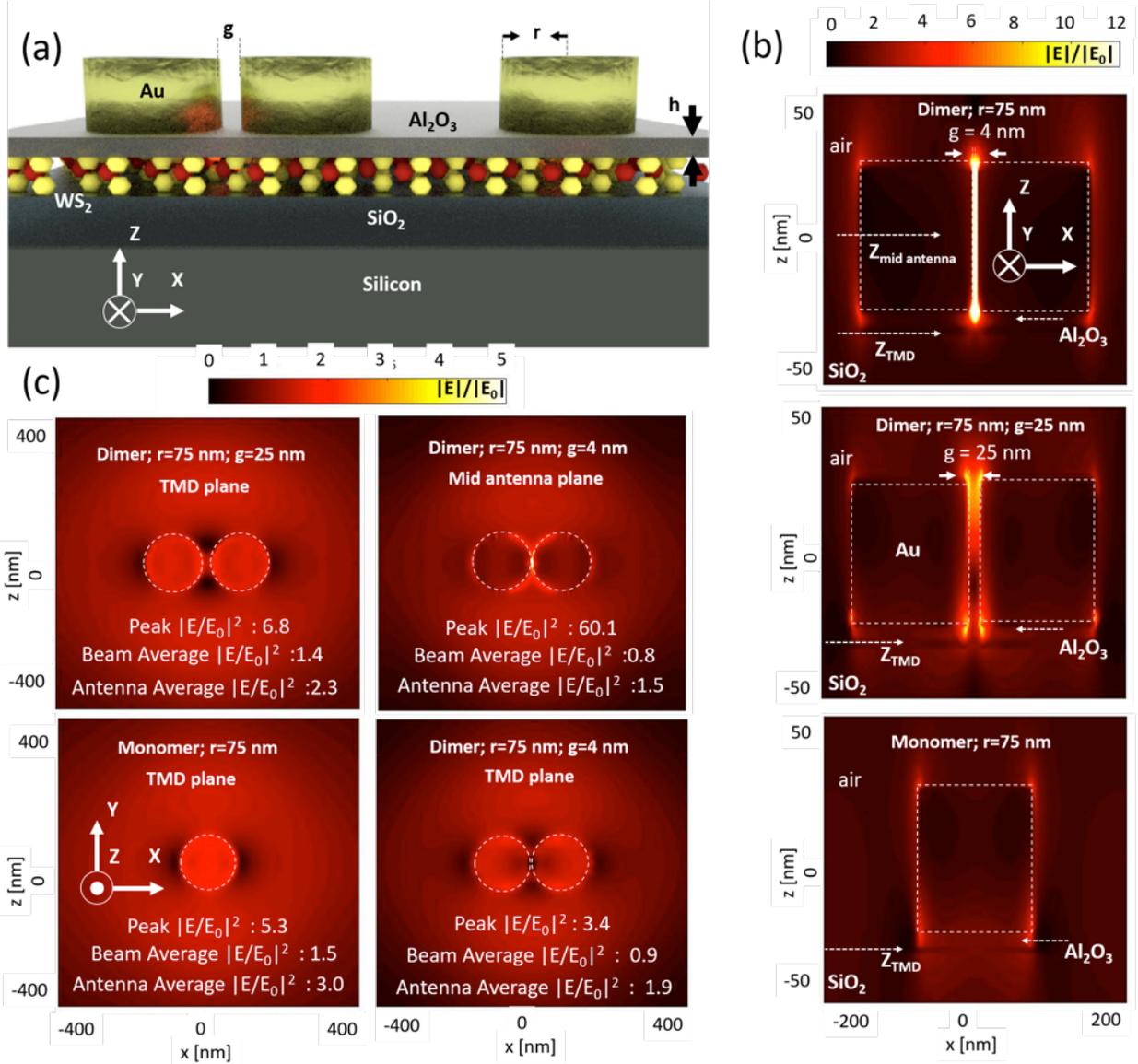

Figure 3. Electric field intensity enhancement (|E|/|E$_0$|) comparison of dimer and monomer antennae showing comparable enhancement at TMD position, which is separated by a spacer layer to avoid quenching. a) Schematic of the purposed optical antenna types for PL enhancement of monolayer TMDs. b) Side view the electric field intensity magnitude enhancement distribution in a 4 nm and 25 nm gapped 75 nm radii dimer antenna and a 75 nm radius monomer antenna, respectively from top to bottom. c) Comparison of electric field intensity enhancement for the same antennae shown in top view and two different z-normal plane position at the midpoint of the antenna and at the z-normal plane where TMD layer is positioned. The maximum value, the averaged value over the area of the beam spot size of the simulation, and the averaged value over the area of geometrical antenna cross-section of (|E|/|E$_0$|)$^2$ is reported for each case.

Regarding the internal photon generation enhancement process of the TMD-cavity system, we focus on the near electric field enhancement and Purcell product in the Eqn. 2. The overall fluorescence enhancement of monolayer WS$_2$ by the plasmonic optical antenna can be expressed as the product of excitation rate enhancement, the spontaneous emission probability enhancement (Purcell effect), and outgoing portion of the spontaneous emission:

$$\frac{I_{out}}{I_0} = \frac{\gamma_{em}}{\gamma_{em}^0} = \frac{\gamma_{exc}}{\gamma_{exc}^0}\frac{\gamma_r}{\gamma_r^0} \qquad (3)$$

where the '0' denotes the intrinsic value, $\gamma_{em}$ is the enhanced and intrinsic fluorescence rate, $\gamma_{exc}$ is the excitation rate at the excitation wavelength of 532 nm, and $\gamma_r$ is the radiative decay rate of the emitter at emission wavelength of 640 nm. Since the emitter is excited optically, the excitation rate enhancement is then proportional to ratio of the squared electric field of the emitter with the optical cavity and without the cavity ($\gamma_{exc}/\gamma_{exc}^0 = |E|^2/|E_0|^2$). Here, care must be taken to describe the physical observable accurately. It is tempting to consider the peak field enhancement of a dimer. This is however not an accurate interpretation of the actual experiment often as well as conducted here. Because the spot size of our pump laser beam is significantly larger than the physical area of the antennae, the excitation and hence the photon generation is not a local, hot-spot like effect, but rather originates from an average across the pump beam dimensions. In order to obtain a) an accurate field enhancement originating from a ~0.8 micrometer large pump diameter, and b) a complete picture of the nature of electric field enhancement distribution due to presence of either monomer or dimer antennae, we calculate the spatially resolved electromagnetic field profile both at the location of the monolayer $WS_2$ as well as at the cross section of the optical antenna for both the excitation and emission wavelengths (Figure 3) (see Methods). The dimer metallic nanoparticles separated by a small gap (4 nm in Figure 3b) supports hybridized plasmon resonances because of the capacitive coupling between the plasmon modes of each nanoparticle. Thus, the often-cited high peak-field enhancement (here 60x) is observed only if a few-atom small point emitter would be positioned precisely inside the gap center (Figure 3c). Even if succeeded (e.g. using dye molecules), the signal could not be collected from this hot-spot only, because even the highest resolution light collectors such as a NSOM averages its signal over an area of hundreds of square nanometers [40]. Since the 2D TMD can neither be placed inside the gap, nor right underneath the metal nanoparticle (to avoid quenching), the only logical position would be to place it below or above the antenna, with a few-nanometer thin oxide spacer (Figure 3a). Thus, when we measure the field enhancement at the position of a TMD flake residing at an optimized 8 nm underneath the metal nanoparticle, the field peak enhancement is only 3.4 fold (Figure 3c). We observe a similar trend for the 100 nm and 200 nm radius dimer antennae (Supporting Information). It is thus not plausible to simply take the peak intensity as the emission process enhancing value. One concludes, that for non-point emitters such as 2D WS$_2$ flake it becomes necessary to define an averaged excitation field enhancement factor such as by integrating the emission over either the physical area of the antenna,

$$\frac{\gamma_{exc}}{\gamma_{exc}^0} = \frac{\iint \frac{|E|^2}{|E_0|^2} ds}{S} \qquad (4)$$

or, more accurately, over the pump beam area, where $S$ is either the geometrical cross section of the antenna or the area of the beam. The aforementioned peak intensity inside a small (4 nm) gap dimer of 60, drops to 1.5 (0.8) when averaged over the antenna (beam) area at the unphysical mid-gap dimer position. For the same antenna, the 3.4x enhancement at the TMD plane underneath the cavity drops to 1.9 (0.9) when averaged over the antenna (beam) (Figure 3c). This shows that no actual excitation rate enhancement is expected for small gap dimers when TMDs are sitting at a quenching-safe distance away from the antenna. Interestingly, when the gap is increased from 4 to 25 nm (as studied here), the average enhancements for the antenna (beam) average increase by 21% (53%) compared to the narrower hot-spot gap dimer case. We also note that the monomer with the same radius offers the highest antenna and beam enhancement (Figure 3c, bottom left corner). With the simpler fabrication of monomers over dimers, these results suggest that monomers are equally well-performing to enhance 2D material PL.

A comparison of the far field (scattering efficiency of disk antenna) and near-field spectra (electric field enhancement) of individual Au nanodiscs (Figure S1, S2, S3) shows that the far-field scattering efficiency peaks at a larger nanoparticle size than the near-field intensity enhancement. Consequently, one should acknowledge that maximum scattering efficiency is not synonymous with highest near-field enhancement as claimed also in previous studies [44-46]. Also, often in optics we seek the highest possible quality factor (Q factor) for highest possible light matter interaction; but in a high-performing antenna we seek the opposite because we desire radiation losses [23, 46].

Our theoretical near-field and far-field study reveals that the observed PL enhancement of monolayer $WS_2$ is due to localized electric field and local density of states at both excitation and emission wavelength enhancing both the excitation and emission rate (Figure 2, 3). Because of reciprocity, optical antennae enhance not only the absorption efficiency (excitation efficiency) upon optical pumping, but also the efficiency of photon generation (Purcell effect). However, since the pump is Stokes-shifted from the emission an antenna could be designed to enhance either process selectively provided Q's are sufficient for this spectral filtering to occur.

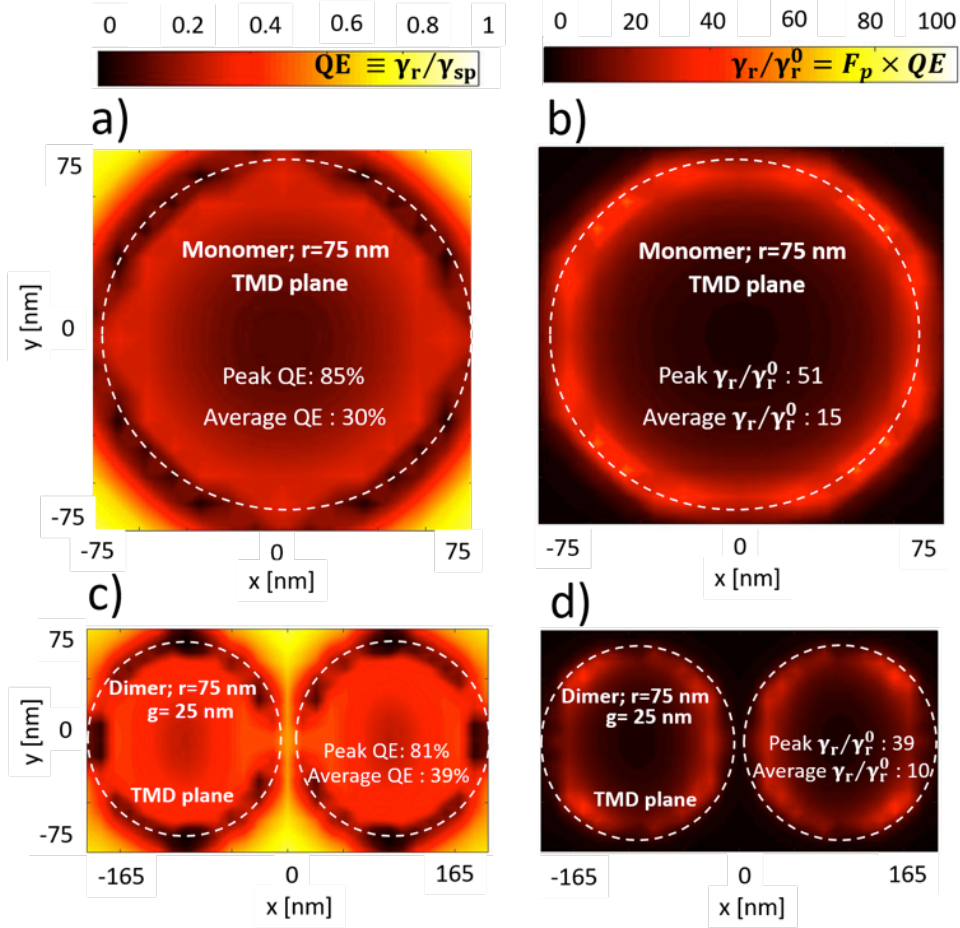

Figure 4. Photon generation rate. Spatial map of the quantum efficiency (a, c) and enhancement in the total radiative rate of $F_p \times QE$ (b, d) for a dipole emitting at a z normal plane corresponding to the position of TMD in the fabricated device. The dashed white lines represent the position of the dimer antennae.

The photon generation rate, here defined as the product between the quantum efficiency and the Purcell factor, is equal to the quantitative radiative decay rate enhancement ($\gamma_r/\gamma_r^0$) (Figure 4 b, d):

$$G = \frac{\gamma_r}{\gamma_r^0} = F_p \times QE = \left(\frac{\gamma_r}{\gamma_{sp}}\right) \times \left(\frac{\gamma_{sp}}{\gamma_r^0}\right) \qquad (5)$$

where $\gamma_{sp}$ is the spontaneous emission rate of the emitter at emission wavelength of ~640 nm, $F_p$ is the Purcell factor, and $QE$ is the quantum efficiency defined here as the portion of spontaneous emission coupled out of the cavity into the free space. We obtain this at the emission wavelength of the monolayer $WS_2$ (~640 nm) positioned under a 75 nm radius monomer or dimer nanocavity compared to an as-grown $WS_2$ flake (Figure 5 a, b) (see methods section for further details). The maximum attained generation rate for a dimer of 75 nm radius and 25 nm gap is ~39 times, and the averaged value over the area underneath the geometrical cross-section of the dimer antenna is about ~10 times the intrinsic radiative rate (Figure 4d). The *QE* of the fluorescence process is estimated by the ratio of the radiated power measured in the far-filed to the total power injected by the emitter (Figure 4 b, d):

$$QE = \frac{\gamma_{rad}}{\gamma_{rad}+\gamma_{loss}+\gamma_{nr}} = \frac{P_{rad}}{P_{rad}+P_{loss}} \qquad (6)$$

However, the decay rate of the excitons to non-radiative channels ($\gamma_{nr}$) such as phonons is not an EM process and is not captured in our FDTD simulation hence is taken as zero. Thus, the average *QE* of the emitter sitting below the dimer cavity is ~39%.

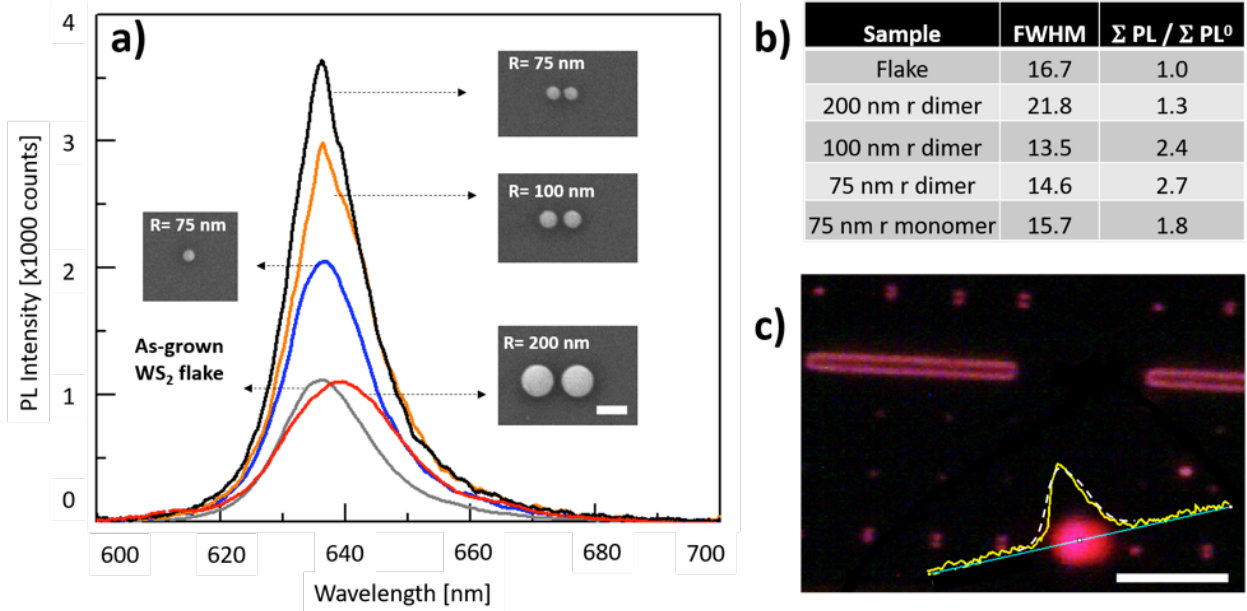

Figure 5. Collected photoluminescence (PL) emission from monolayer $WS_2$ is enhanced when it is placed under a plasmonic monomer or dimer antenna cavity with a resonance close to that of the emission wavelength. a) PL intensity of CVD-grown monolayer $WS_2$ before and after fabrication of 4 different optical antennae. Insets are SEM images of each type of optical cavity; the scale bar is 400 nm. (b) Full Width Half Maxima (FWHM) and enhancement of integrated PL emission from the as-grown sample and for each optical antenna. c) Optical image of monolayer WS$_2$ emission (637 nm) under the 75 nm radius dimer antenna. The scale bar is 5 microns and 90% of the Gaussian beam spot power spot is within ~800 nm length.

To quantify our numerical estimate of PL enhancement factor, we use all the numerical values obtained from Equation 2 throughout this paper. Based on these results, we expect an averaged PL enhancement over the beam spot (geometrical cross section) for the 75 nm radius monomer and dimer antenna-cavities as 2.1 (30.0) and 3.8 (23.0) fold respectively. To compare the theoretical values with experimental results, it is customary to compare experimental measurement results directly with the simulation results averaged over the beam spot of ~0.8 μm (Figure. 5b). In our measurements, we fit the beam intensity profile to a Gaussian function and take the FWHM as the beam spot size (Figure. 5c). These experimental results closely relate to our theoretical calculation, but show slightly lower values. Another common way of expressing experimental results is to define the normalized experimental enhancement factor ($< EF_{exp} >$) as:

$$< EF_{exp} > = \frac{I_{cav}}{I_0} \Big/ \frac{S_{cav}}{S_0} \qquad (7)$$

where $S_{cav}$ is the area of a cavity, $S_0$ is the area of the Gaussian excitation beam, $I_{cav}$ is the PL intensity from the cavity, and $I_0$ is the PL intensity from the as-grown WS$_2$ flake; resulting in an $< EF_{exp} >$ of 50.9 and 39.7 fold. This value is clearly an overestimation of the experimental result, because per our simulation results the electric field intensity and Purcell factor distribution directly under nanoantenna are significantly larger than the area around the antenna. Thus, here we suggest that for a more accurate comparison of PL enhancements the average effect of over the entire beam spot should be taken into consideration instead.

The shape of the PL spectrum on the cavity is conceptually similar to that of the control sample. However, we observe a spectral narrowing in FWHM value from 16.7 nm to 15.7, 13.5, and 14.6 nm respectively for the cases of 75 nm radius monomer disc antenna, and 75 nm and 100 nm radii dimer antennae, respectively. On the other hand, the PL spectra of the 200-nm radius dimer antenna appears to be red shifted by 3.5 nm relative to the intrinsic emission spectrum of as-grown $WS_2$ flake, and we observe a band broadening to 21.8 nm. This redshift appears due to convolution of the PL emission spectra of monolayer $WS_2$ with the fundamental resonance of the 200-nm radius dimer cavity scattering spectrum for this nanocavity occurring at higher wavelength (Supporting Information). Narrowing of the emission response occurs for the case of 75 nm and 100 nm radii dimers when resonance of the dimer cavities matches with that of the emission peak of $WS_2$, resulting in higher quality factor and sharper spectral response (Supporting Information). From Wheeler's limit, it is expected to find higher quality factors for smaller antennae [47]. Such narrowing of the spectral response due to smaller cavity radius is evident from our experimental results summarized in Figure 5, which suggests that the quality factor of nanoantenna increases with reducing dimensions of each disc.

In conclusion, we have demonstrated that optical nanoantenna can be used to control the emitting properties of monolayer TMDs. This control was achieved using two types of the metallic cavities (monomer vs dimer) in four different dimensions. These emission dynamics were also supported by numerical calculations. In particular, we have demonstrated that the fluorescent enhancement of 2D materials unlike quantum dots is an areal average effect. We have also observed band narrowing of the emission response when resonance of the cavity corresponds to emission wavelength of monolayer $WS_2$. Both monomer and dimer nanoantenna architecture are scalable to emission resonances of other members of the TMD family as well such as $MoTe_2$ which emits at telecommunications wavelengths in the near infrared. The demonstrated nano-antenna controlled emission from a monolayer $WS_2$ flake could open a pathway to visible light sources based on lithographically fabricated nano-antennae supporting a variety of opto-electronic applications [48-53].

**METHODS**

Growth method

Single- and Multi-layer Tungsten Disulfide ($WS_2$) was grown via ambient-pressure chemical vapor deposition (CVD) utilizing a tube furnace. The process is a variation of our previously published work on transition metal dichalcogenide materials [54-56]. The reagents were ammonia meta tungstate (AMT) and elemental sulfur. The process starts by spin-coating a 3.1mmol aqueous solution of AMT onto a $SiO_2$/Si substrate. The resultant residue serves as the tungsten source for our CVD growth. We place a target substrate directly face-down onto this source substrate and insert this stack into the growth region at the center of the process tube in our tube furnace. An alumina boat with containing elemental sulfur is placed inside the process tube far enough upstream of the furnace heating coil so that it just fully melts when the furnace center reaches the peak growth temperature of 850°C.

The temperature ramp for $WS_2$ growth commences after a nitrogen purge of the process tube for 15 minutes at 5 SCFH. After that we ramp the furnace to 500°C so as to decompose the AMT into tungsten trioxide releasing water and ammonia Vapor. After 20 minutes for this reaction to complete, we ramp up to the growth temperature of 850°C and remain there for 15 minutes. Subsequently, the furnace is allowed to cool naturally to 200°C before the process tube is opened to air and the target substrate retrieved.

Fabrication method

We fabricate monomer and dimer disk optical antennae with various diameters and gaps onto the large area $WS_2$ film by conventional electron beam lithography followed by electron beam evaporator gold deposition, and finally a lift-off process in PG remover. Here, lithographical fabrication of the antennae give us control over dimensions and position of particles, allowing us to tune plasmonic resonance of each antenna element.

After deposition of $WS_2$ thin layers, a very thin layer of aluminum oxide was deposited on the sample as a spacer between the emitter and plasmonic antennae. The alumina spacer layers were deposited on top of the $WS_2$ samples by iterative deposition of thin (<2nm) aluminum using electron beam evaporation, followed by natural oxidation to $Al_2O_3$ under ambient conditions. The thickness of the film is monitored through quartz crystals during the deposition, and estimated by an ellipsometry after the oxidation process. The $Al_2O_3$ film functions as the spacer layer between $WS_2$ and gold optical antennae. Next, the nanoantenna were prepared and arranged in a square array with the center-to-center distance between the antennae designed to be as large as 4 microns to prevent interference of absorbing cross section of individual antenna (Supporting Information). To increase the stability of the fabricated nanostructures during lift-off, a thin layer of chromium (typically < 2 nm) is used as the adhesion layer. Only larger patches of metal survive lift-off without such precautions.

<u>Raman and Photoluminescence Spectroscopy</u>

Raman and PL spectroscopy were performed using a custom-built spectrometer equipped with a 532-nm excitation line and CCD detector (Supporting Information). The nonlinear emission spectra were acquired in reflection geometry to a spectrometer including a 532-nm notch filter to reject the pump wavelengths. The spectrometer acquisition parameters were held constant and set to ensure high signal-to-noise ratio for the weakest signal. All measurements were performed in air at ~25 Celsius and atmospheric pressure.

Reflectance Spectroscopy

Reflectance measurements were performed using a 20W halogen light source by measuring the difference in reflected intensity from the $WS_2/SiO_2$ and bare silicon wafer substrate and normalizing this to the substrate-reflected intensity. For optical microscopy, we used white light and a 100x objective lens. All measurements were performed in air at ~25 Celsius and atmospheric pressure.

FDTD calculation method

Reflectance, in-plane and out-of-plane electric field intensities, Purcell factor, scattering, and absorption spectra were calculated. Optical constants of $WS_2$ were obtained from Liu et al [57]. The refractive index of silica, silicon, alumina, and gold were used directly from Palik, and Johnson and Christy et al [58, 59].

We used Lumerical three-dimensional finite-difference time-domain (FDTD) solver (Figure 3a, b) for all Maxwell equation calculations. For scattering effect calculations, we used a total field scattered field (TFSF) method. The plane wave was launched normally from the top of monomer or dimer antennae. Two power-flow monitors (six detectors each) were positioned inside and outside the TFSF source, surrounding the antenna, to measure the absorption and scattering cross sections, in order. The power flow analysis measures the net absorbed power and scattered power from particle (Figure 3). In case of the monomer antenna, we sweep the particle radius from 25 nm up to 200 nm. For Purcell factor calculations, a dipole source was used. In all other problems, a normal incident broadband plane wave was implemented. To ensure that scattered light does not return to the simulation region a Perfect Matching Layer (PML) was applied as boundary condition in all six directions.

For Purcell factor and Quantum efficiency calculations, we mapped spatial position of a dipole source at the z-normal plane corresponding to position of the 2D TMD sheet. We calculated the radiated power by a set of field power monitors as one transmission box surrounding only the dipole source and another transmission box far away from source surrounding both the emitter and cavity. The expression deployed in this method to evaluate Purcell factor is $F_p$ = Emitted-Power ($f$) / Source-Power ($f$), where $f$ is the optical frequency. The Source-Power returns the power that the dipole would radiate in a homogeneous material. The latter is the power transmitted out of a boxed area surrounding the dipole source due to the dispersive materials used in the cavity.

*Supporting Information Available*: [Description of near field intensity distribution, scattering efficiency, absorption loss map, and quality factor for all dimer antennae; additional Purcell factor results for a narrow gap dime antenna; nanoantenna fabrication steps and SEM images; PL measurement setup; DFT calculation results] The Supporting Information is available free of charge on the ACS Publications website at DOI: 10.1021/acsphotonics.XXXXXXX.

**ACKNOWLEDGMENT**

We thank the National Science Foundation and the Materials Genome Initiative for support under the award number NSF DMREF 14363300/1455050 and NSF EAPSI 1613966. V.S. is supported by Air Force Office of Scientific Research Young Investigator Program under grant FA9550-14-1-0215 and FA9550-14-1-0378.